\documentclass[preprint,12pt]{elsarticle}
\journal{Computer Physics Communications}

\usepackage[T1]{fontenc}
\usepackage{framed}
\usepackage[utf8]{inputenc}
\usepackage{url}
\usepackage{graphicx}
\usepackage{amsmath}
\usepackage{amssymb}
\usepackage{hyperref}
\usepackage{longtable}
\usepackage{color}
\usepackage{algorithm}
\usepackage{algorithmicx}
\usepackage{amssymb}
\usepackage{mathrsfs}
\usepackage{subcaption}
\usepackage{array}
\usepackage{booktabs}
\usepackage{multirow}
\graphicspath{{figures/}}

\newcommand{\ui}{\mathrm{i}}
\newcommand{\ud}{\mathrm{d}}

\begin{document}
\begin{frontmatter}

\title{{\tt AMFlow} 2.0: significant algorithmic and software improvements for Feynman integral evaluation}

\author[a]{Rui-Jun Huang}
\ead{huangrj2215@pku.edu.cn}

\author[b,c]{Xiao Liu\corref{author}}
\ead{xiao.liu@ucas.ac.cn}

\author[a,d]{Yan-Qing Ma\corref{author}}
\ead{yqma@pku.edu.cn}

\cortext[author]{Corresponding author.}

\address[a]{School of Physics, Peking University, Beijing 100871, China}
\address[b]{School of Physical Sciences, University of Chinese Academy of Sciences, Beijing 100049, China}
\address[c]{Theoretical Physics Department, CERN, 1211 Geneva, Switzerland}
\address[d]{Center for High Energy Physics, Peking University, Beijing 100871, China}

\begin{abstract}
We present significant improvements to the {\tt AMFlow} package for the numerical computation of dimensionally regularized Feynman integrals. Several new features are introduced to reduce computational cost, including an alternative recursion mode, a high-performance differential equation solver, support for state-of-the-art integration-by-parts reducers and other useful improvements. We benchmark the new version on a three-loop five-point topology and find that both the symbolic and numerical performance are significantly improved.
\end{abstract}


\end{frontmatter}

\newpage

\textbf{PROGRAM SUMMARY}

\vspace{1cm}

\begin{small}
	\noindent
	{\em Program title:} {\tt AMFlow} 2.0\\
	{\em Developer's repository link:} \url{https://gitlab.com/multiloop-pku/amflow}\\
	{\em Licensing provisions:} MIT\\
	{\em Programming language:} {\tt Wolfram Mathematica}, {\tt C++}\\
    {\em Journal Reference of previous version:}

    Xiao Liu and Yan-Qing Ma, AMFlow: A Mathematica package for Feynman integrals computation via auxiliary mass flow, \href{https://www.sciencedirect.com/science/article/pii/S0010465522002843?via\%3Dihub}{Comput.Phys.Commun. 283 (2023) 108565}, \href{https://arxiv.org/abs/2201.11669}{e-Print: 2201.11669}\\
    {\em Does the new version supersede the previous version?:} Yes. \\
    {\em Reasons for the new version:} Algorithmic and software improvements yielding significant performance gains on both the symbolic and numerical sides.\\
    {\em Summary of revisions:} The new features include an alternative recursion mode, a high-performance differential equation solver, support for state-of-the-art integration-by-parts reducers, basis refinement and several other enhancements.\\
    {\em Nature of problem:} High-precision numerical evaluation of dimensionally regularized Feynman integrals.\\
	{\em Solution method:} Recursive simplification of Feynman integrals combined with high-performance integration-by-parts reducers and differential equation solver.\\
	{\em External routines/libraries used:} {\tt Wolfram Mathematica} [1], {\tt FiniteFlow} [2], {\tt LiteRed} [3], {\tt Kira} [4], {\tt FIRE} [5], {\tt Blade} [6], {\tt Ratracer} [7], {\tt MPSolve} [8], {\tt GMP} [9], {\tt MPFR} [10], {\tt MPC} [11], {\tt Boost} [12], {\tt yaml-cpp} [13]\\
	{\em References:}
	{\\} [1] \url{http://www.wolfram.com/mathematica}, commercial algebraic software;
	{\\} [2] \url{https://github.com/peraro/finiteflow}, open source;
	{\\} [3] \url{http://www.inp.nsk.su/~lee/programs/LiteRed}, open source;
	{\\} [4] \url{https://gitlab.com/kira-pyred/kira}, open source;
	{\\} [5] \url{https://gitlab.srcc.msu.ru/feynmanintegrals/fire}, open source;
	{\\} [6] \url{https://gitee.com/multiloop-pku/blade}, open source;
	{\\} [7] \url{https://github.com/magv/ratracer}, open source;
	{\\} [8] \url{https://github.com/robol/MPSolve}, open source;
	{\\} [9] \url{https://gmplib.org/}, open source;
	{\\} [10] \url{https://www.mpfr.org/}, open source;
	{\\} [11] \url{https://www.multiprecision.org/mpc/}, open source;
	{\\} [12] \url{https://www.boost.org/}, open source;
	{\\} [13] \url{https://github.com/jbeder/yaml-cpp}, open source.
\end{small}
\newpage

\section{Introduction}\label{sec:intro}

The Standard Model (SM) of particle physics is among the most precisely tested theories in physics. Its predictive power fundamentally relies on perturbative quantum field theory, where physical observables are expressed as asymptotic series in coupling constants. In this context, reliable higher-order perturbative calculations are not merely theoretical refinements but are essential for uncovering small deviations from the SM and constraining possible new physics.

The computation of Feynman integrals is crucial for such perturbative calculations. Schematically, integrals are first reduced to a smaller set of master integrals using integration-by-parts (IBP) reduction ~\cite{Chetyrkin:1981qh, Laporta:2000dsw, Gluza:2010ws, Larsen:2015ped, vonManteuffel:2014ixa, Peraro:2016wsq, Mastrolia:2018uzb, Driesse:2024xad, Bern:2024adl}; the master integrals are then evaluated via various methods~\cite{Binoth:2000ps, Boos:1990rg, Smirnov:1999gc, Tausk:1999vh, Kotikov:1990kg, Henn:2013pwa, Boughezal:2007ny, Czakon:2008zk, Borowka:2017idc, Hidding:2020ytt, Armadillo:2022ugh}, among which the auxiliary mass flow method~\cite{Liu:2017jxz, Liu:2021wks, Liu:2022mfb} provides a systematic and efficient solution. By introducing an auxiliary mass parameter $\eta$ into some of the denominators of a dimensionally regularized Feynman integral family,
\begin{align}
I(\eta)=\int\prod_{i=1}^{L}\frac{\ud^{D}\ell_i}{\ui\pi^{D/2}}
\frac{\mathcal{D}_{k+1}^{-\nu_{k+1}}\cdots \mathcal{D}_N^{-\nu_N}}{(\mathcal{D}_1 -q_1\eta)^{\nu_{1}}\cdots (\mathcal{D}_k-q_k\eta)^{\nu_{k}}}, \quad q_j \in \{0, 1\},
\end{align}
we can simplify the corresponding integrals by expanding around $\eta=\infty$ and obtain the result at $\eta=\ui 0^-$ using differential equation methods. Several public codes implementing this method are available. The {\tt AMFlow} package ~\cite{Liu:2022chg} was the first proof-of-concept implementation and supports extensions to cut integrals ~\cite{Liu:2020kpc} and integrals with linear propagators ~\cite{Liu:2022tji}, whereas the {\tt TTH}~\cite{Buccioni:2023okz} and {\tt DCT}~\cite{Huang:2024qan} packages focus primarily on one-loop applications, and the {\tt LINE}~\cite{Prisco:2025wqs} package emphasizes computational performance through a low-level implementation.

However, as the complexity of the problems increases, the application of the auxiliary mass flow method becomes increasingly challenging. These challenges arise from two main aspects: (1) on the symbolic side, target integrals must be reduced and differential equations for master integrals must be constructed using IBP reduction, whose efficiency is constrained both by the structure of the auxiliary mass flow recursion and by the performance of available IBP tools~\cite{vonManteuffel:2012np, Lee:2013mka, Peraro:2019svx, Wu:2023upw, Wu:2025aeg, Guan:2024byi, Smirnov:2025prc, Maierhofer:2017gsa, Klappert:2020nbg, Lange:2025fba}; (2) on the numerical side, the differential equations must be solved using high-precision arithmetic, whose efficiency depends strongly on the underlying implementation.

In this paper, we present {\tt AMFlow} 2.0, a major update of the {\tt AMFlow} package, which is available at
\begin{align}
	\text{\url{https://gitlab.com/multiloop-pku/amflow}}.
\end{align}
This update introduces a number of new features aimed at reducing both aspects of the aforementioned computational costs. These features, including an alternative recursion mode, a {\tt C++} implementation of the differential equation solver, improved IBP interfaces, basis refinement, and several other enhancements, are described in detail in Section~\ref{sec:analytical}. We also provide benchmarks in Section~\ref{sec:tests}, followed by conclusions in Section~\ref{sec:outlook}.

\section{New features}\label{sec:analytical}

\subsection{Alternative recursion mode}\label{sec:feyn_para}

Besides the auxiliary mass flow method, other methods that can recursively simplify integral topologies also exist~\cite{Hidding:2022ycg, Chen:2023hmk, Chen:2024xwt, Chen:2025paq}. These methods are able to reduce the number of propagators before entering the next step of the recursion and typically do not lead to a proliferation of master integrals. While evaluations in the physical region using these methods are generally less stable, they can serve as a useful supplement to the auxiliary mass flow method. We therefore implement one representative method, the Hidding--Usovitsch approach~\cite{Hidding:2022ycg}, in \texttt{AMFlow} 2.0 as an alternative recursion mode.

This approach works as follows. For target integrals in a given integral family, we first introduce a Feynman parameter $x$ to combine a pair of propagators:
\begin{equation} \label{eq:FT}
\frac{1}{\mathcal{D}_i^{\nu_i} \mathcal{D}_j^{\nu_j}}
= \frac{\Gamma(\nu_i + \nu_j)}{\Gamma(\nu_i)\Gamma(\nu_j)}
\int_0^1 \ud x\, \frac{x^{\nu_i-1} (1-x)^{\nu_j-1}}{\left(x\mathcal{D}_i+(1-x)\mathcal{D}_j\right)^{\nu_i+\nu_j}}.
\end{equation}
Every integral containing these two propagators in the original family can then be expressed as an integral over $x$, whose integrand can in turn be interpreted as a Feynman integral in the \textit{combined family} with one fewer propagator. Next, we construct a system of differential equations with respect to $x$ for the master integrals in the combined family and compute the general solution for these master integrals, and hence for the integrand in Eq.~\eqref{eq:FT}. The boundary constants can then be fixed by evaluating the master integrals at a generic point $x=x_0$, which can itself be achieved by applying the above procedure recursively. In this way, we simplify the problem until we arrive at a single-propagator topology, whose evaluation is straightforward. After the boundary constants have been determined, the integral in Eq.~\eqref{eq:FT} can be evaluated to obtain the original result.

To use this approach in {\tt AMFlow} 2.0, the user can simply set the option \texttt{"RecursionMode"} to \texttt{"FT"} (the default value is \texttt{"AMF"}, corresponding to the traditional auxiliary mass flow approach):
\par
\texttt{SetAMFOptions["RecursionMode" $\to$ "FT"];}\\
where {\tt "FT"} stands for Feynman's Trick. We emphasize that, since this mode requires integration over the Feynman parameter from 0 to 1, it is necessary to determine the correct contour deformation whenever a physical singularity appears along the integration path. Unfortunately, only heuristic rules for such contour deformations are currently available, and a systematic prescription is still lacking~\cite{Hidding:2022ycg}. Therefore, in our implementation we simply deform the contour randomly, and as a result only evaluations in the Euclidean region are guaranteed to be correct. We leave the treatment of physical kinematics for future study.\footnote{See Refs.~\cite{Chen:2023hmk, Chen:2024xwt, Chen:2025paq} for some potential treatments in Feynman-parameter space.}

\subsection{High-performance differential equation solver}

We have developed a high-performance differential equation solver in {\tt C++}, which now serves as the default solver, replacing the previous implementation written in \texttt{Mathematica}. The {\tt C++} solver covers all previous functionality while also providing numerous additional features required by the \texttt{"FT"} recursion mode introduced in Section~\ref{sec:feyn_para}, as well as other potential applications.

For seamless integration, we have designed a {\tt WSTP} (Wolfram Symbolic Transfer Protocol) link that allows the {\tt C++} solver to be accessed from the \texttt{Mathematica} frontend. In this framework, \texttt{Mathematica} handles problem setup, symbolic analysis, and data processing, while the {\tt C++} backend performs the intensive numerical computations. Through this link, we are able to keep the structure of the \texttt{Mathematica} scripts unchanged while taking full advantage of the performance of {\tt C++}.

In addition to the {\tt WSTP} link, we have also provided a standalone, {\tt YAML}-configuration-driven command-line executable named {\tt desolver}. Since this executable does not depend on \texttt{Mathematica}, it can be used outside the context of \texttt{AMFlow} and should be regarded as a general-purpose differential equation solver.

For installation and basic usage of the {\tt C++} solver, we refer the reader to {\tt README.md} and {\tt examples/differential\_equation\_solver\_cpp}.

\subsection{Improved IBP interfaces}\label{sec:ibp_interface}

The computational cost of IBP reductions depends not only on the performance of the IBP reducers themselves, but also on the way in which \texttt{AMFlow} interacts with them. In \texttt{AMFlow} 2.0, we have updated the interfaces to support state-of-the-art IBP reducers, including {\tt Blade}~\cite{Guan:2024byi}, {\tt FIRE} 7~\cite{Smirnov:2025prc}, and {\tt Kira} 3~\cite{Lange:2025fba}, and have optimized the algorithms used to interface with them, thereby improving the performance of our computational framework. For {\tt Kira} 3, we further introduce {\tt Ratracer}~\cite{Magerya:2022hvj} as an alternative linear solver, which can be activated by\par
\texttt{SetReducerOptions["ReductionMode" $\to$ "Ratracer"];}\\
While our interfaces do not yet exploit the full capabilities of these tools, even their basic functionality now enables problems that were previously beyond the reach of {\tt AMFlow} 1.2. See Section~\ref{sec:tests} for more details.

In addition to the aforementioned reducers, several other reducers are also very promising in the context of {\tt AMFlow}, such as {\tt NeatIBP}~\cite{Wu:2023upw,Wu:2025aeg} and {\tt CALICO}~\cite{Bertolini:2025zud} in combination with {\tt FiniteFlow}~\cite{Peraro:2019svx}. We leave the integration of these tools to future work.

\subsection{Refined master integrals}

It is well known that one can always choose a basis of master integrals such that the denominators of the reduction coefficients factorize into a product of an $\epsilon$-dependent part and a kinematics-dependent part~\cite{Usovitsch:2020jrk, Smirnov:2020quc}. For complicated problems, such a factorized basis can typically reduce the size of IBP reduction tables. Furthermore, in the context of \texttt{AMFlow}, it can also improve numerical stability because \texttt{AMFlow} uses $\epsilon$-sampling to reconstruct the final $\epsilon$-expansions, where $\epsilon$ may take very small numerical values.

We therefore implement the algorithm of Refs.~\cite{Usovitsch:2020jrk, Smirnov:2020quc} in \texttt{AMFlow} 2.0 to transform the default basis into a factorized basis before constructing differential equations at each step of the recursion. This feature can be enabled by
\par
\texttt{SetAMFOptions["RefineBasis" -> True];}\\
However, it is worth noting that this option is disabled by default because the basis-transformation algorithm may, in some cases, increase the computational burden on the symbolic side. On the one hand, the algorithm requires an additional reduction. On the other hand, ensuring the reduction to the refined basis may require a larger IBP system, which can further increase the computational cost. Therefore, we recommend enabling this option only when the numerical cost dominates the symbolic cost, for example when results with $\mathcal{O}(100)$ digits of precision are required.

\subsection{Other improvements}\label{sec:others}

We have also added several other useful functionalities to improve performance and stability. A caching mechanism has been introduced to enable interrupted computations to be resumed, which can be activated by
\par
\texttt{SetAMFOptions["UseCache" -> True];}\\
When this option is enabled, all computations on the symbolic side, such as the construction of differential equations and the analysis of boundary integrals, will be reused whenever possible.

Another noteworthy feature is the \texttt{"SkipReduction"} option. Sometimes, the input target integrals already form a basis, or a subset of a basis, of master integrals, and the reduction to master integrals performed by \texttt{AMFlow} is therefore unnecessary. In such cases, we recommend enabling this option by
\par
\texttt{SetAMFOptions["SkipReduction" -> True];}\\
to skip that reduction step.

For other new features and changes, we refer the reader to \texttt{CHANGELOG.md}.

\section{Benchmark}\label{sec:tests}

Massless three-loop five-point integrals represent a frontier in perturbative quantum field theory. Analytic results for planar topologies have been obtained in Refs.~\cite{Chicherin:2025mvc, Liu:2024ont} using the canonical differential equation method~\cite{Kotikov:1990kg, Henn:2013pwa}, where cross-checks with {\tt AMFlow} 1.2 were reported to be prohibitively expensive due to the cost of the required IBP reductions.

\begin{figure}
\centering
\includegraphics[width=0.6\textwidth]{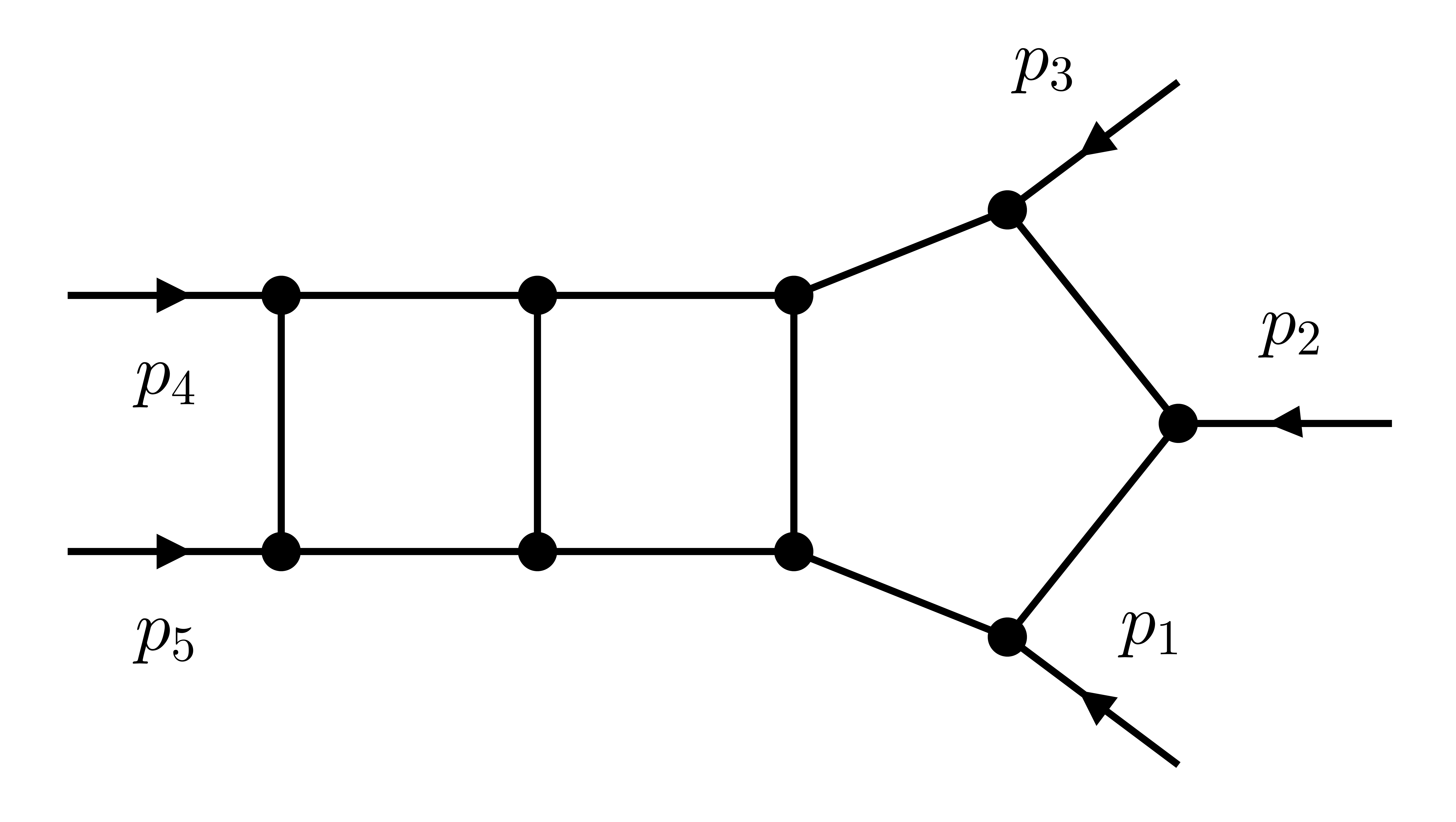}
\caption{The {\tt PBB} integral family.}
\label{fig:pbb}
\end{figure}

We therefore benchmark the new version of {\tt AMFlow} on one of the three-loop five-point topologies, namely {\tt PBB} from Ref.~\cite{Liu:2024ont}, whose Feynman diagram is shown in Fig.~\ref{fig:pbb}. In order to comprehensively assess the performance of the package, we evaluate the 316 master integrals using different combinations of recursion modes ({\tt "AMF"}, {\tt "FT"}), IBP reducers ({\tt Blade}, {\tt Kira} 3.1 with {\tt Ratracer}), and differential equation solvers ({\tt C++}, {\tt Mathematica}).

Since the {\tt "FT"} mode is currently guaranteed only in the Euclidean region, we choose a Euclidean phase-space point defined by
\begin{equation}
\vec{s} = \{s_{12},s_{23},s_{34},s_{45},s_{51}\} = \left\{-\frac{36}{29}, -\frac{32}{23}, -\frac{45}{31}, -\frac{33}{19}, -\frac{25}{18}\right\},
\label{eq:point}
\end{equation}
where $s_{ij} = (p_i + p_j)^2$. We use the command
\par
{\tt SolveIntegrals[masters, 20, 6];}\\
to obtain the $\epsilon$ expansions for the master integrals up to finite terms with 20 correct digits.

Tables~\ref{tab:amflow_pbb} and~\ref{tab:ft_pbb} summarize the computational costs for the {\tt PBB} family using the {\tt "AMF"} and {\tt "FT"} modes, respectively. For each mode, we test all combinations of reducers and solvers. The symbolic cost $t_\text{sym}$ includes all IBP reductions and the construction of differential equations, while the numerical cost $t_\text{num}$ corresponds to the numerical stage, including differential-equation solving and the manipulation of (asymptotic) expansions. The total runtime is given by $t_\text{tot}=t_\text{sym}+t_\text{num}$. All results have been validated against reference values obtained with {\tt DiffExp}~\cite{Hidding:2020ytt} using the boundary conditions provided in Ref.~\cite{Chicherin:2025mvc}.

\begin{table}[htbp]
	\centering
    \begin{tabular}{|c|c|c|c|c|}
    \hline
    \multicolumn{5}{|c|}{Recursion mode: {\tt "AMF"}}\\
        \hline
       IBP reducer & $t_\text{sym}$ & Differential equation solver & $t_\text{num}$ & $t_\text{tot}$ \\
        \hline
        \multirow{2}{*}{\tt Blade} & \multirow{2}{*}{608.0} & {\tt C++} & 97.4    & 705.4       \\
          &                        & {\tt Mathematica} & 268.3    & 876.3       \\
        \cline{1-2}
        \hline
        \multirow{2}{*}{{\tt Kira 3.1} $+$ {\tt Ratracer}}   & \multirow{2}{*}{496.5} & {\tt C++} & 103.2    & 599.7       \\
           &                        & {\tt Mathematica} & 331.0   & 827.5       \\
        \hline
    \end{tabular}
	\caption{Computational costs for the {\tt PBB} family using the {\tt "AMF"} recursion mode. All times are reported in CPU hours.}
	\label{tab:amflow_pbb}
\end{table}

\begin{table}[htbp]
	\centering
    \begin{tabular}{|c|c|c|c|c|c|}
    \hline
    \multicolumn{5}{|c|}{Recursion mode: {\tt "FT"}}\\
        \hline
        IBP reducer & $t_\text{sym}$ & Differential equation solver & $t_\text{num}$ & $t_\text{tot}$  \\
        \hline
        \multirow{2}{*}{\tt Blade} & \multirow{2}{*}{173.9} & {\tt C++} & 28.1    & 202.0       \\
          &                        & {\tt Mathematica} & 699.0    & 872.9       \\
        \cline{1-2}
        \hline
        \multirow{2}{*}{{\tt Kira 3.1} $+$ {\tt Ratracer}}  & \multirow{2}{*}{133.7} & {\tt C++} & 40.6    & 174.3       \\
           &                        & {\tt Mathematica} & 811.7    &  945.4      \\
        \hline
    \end{tabular}
	\caption{Computational costs for the {\tt PBB} family using the {\tt "FT"} recursion mode. All times are reported in CPU hours.}
	\label{tab:ft_pbb}
\end{table}

It can be seen from the tables that the symbolic cost of the {\tt "FT"} mode is significantly lower than that of the {\tt "AMF"} mode, as expected. This is because the number of master integrals at the first step of the recursion is much smaller: 150 for {\tt "FT"} compared to 521 for {\tt "AMF"}. For each recursion mode, the {\tt C++} solver also exhibits better performance than the {\tt Mathematica} solver, as expected.

 An interesting observation is that, when using the {\tt Mathematica} solver, the numerical cost of the {\tt "FT"} mode is much larger than that of the {\tt "AMF"} mode and even exceeds the symbolic cost. This behavior can be attributed to two main reasons. First, the algorithm used by the {\tt Mathematica} solver to solve differential equations arising in the {\tt "FT"} mode is not yet fully optimized. Second, computing the $x$ expansions of the integrand appearing in Eq.~\eqref{eq:FT} is itself a nontrivial task and can consume a considerable amount of time. Fortunately, the {\tt C++} solver benefits from both an optimized implementation of the relevant algorithms and the superior performance of a compiled language. As a result, it provides an overall speedup of roughly a factor of 20, and the numerical cost no longer dominates the computation.

\section{Summary and outlook}\label{sec:outlook}

We have presented {\tt AMFlow} 2.0, a major update of the {\tt AMFlow} package, which is available at
\begin{align}
\text{\url{https://gitlab.com/multiloop-pku/amflow}}.
\end{align}
This update introduces a number of new features, including the {\tt "FT"} recursion mode, a {\tt C++} differential equation solver, support for state-of-the-art IBP tools, and several other improvements. As demonstrated by our benchmark on a representative three-loop five-point topology, these developments can significantly reduce the computational costs on both the symbolic and numerical sides.

The {\tt "FT"} mode benefits from a smaller number of master integrals during the recursion, leading to a substantial reduction in the symbolic cost. The trade-off is that evaluations in physical regions are currently less stable. A systematic prescription for contour deformation is still lacking, and we plan to investigate this issue in future work.

As shown in Section~\ref{sec:tests}, when the required numerical precision is moderate, the symbolic cost can still dominate the overall computation. Therefore, the incorporation of new ideas and tools from IBP reduction~\cite{Wu:2025aeg, Guan:2023avw, Feng:2025leo, Feng:2026imq, Huang:2026xnq, Smith:2025xes, Liu:2025udl, delaCruz:2026mas, vonGersdorff:2026zco, vonHippel:2025okr, Song:2025pwy, Zeng:2025xbh, Berman:2026jgq, Fontana:2023amt, Chestnov:2024nbj, Bertolini:2025zud}, as well as improved strategies for selecting master integrals~\cite{e-collaboration:2025frv, Bree:2025tug}, remains an important direction for future development.

Another important direction is to achieve fast evaluations across the full phase space. This will require combining the current framework with efficient techniques for analytic continuation between different kinematic regions (see Refs.~\cite{PetitRosas:2025xhm, Badger:2025ljy, Liu:2026hdp, Baur:2026zlw, Liu:2026cpf, Czakon:2026tog, Abreu:2026vxw} for recent developments). We leave such an integration for future study.

\section*{Acknowledgements}
We thank Samuel Abreu, Gaia Fontana, Xiang Li, Vitaly Magerya, Pier Francesco Monni, Johann Usovitsch, Xing Wang and Yang Zhang for many useful discussions. We thank Yongqun Xu for assistance in comparing their results of three-loop five-point integrals with ours. Rui-Jun Huang and Yan-Qing Ma are supported by the National Natural Science Foundation of China (NSFC) through Grant No. 12325503. Xiao Liu is supported by the Fundamental Research Funds for the Central Universities and the European Union (ERC, grant agreement No. 101044599).

\bibliographystyle{utphysMa}
\providecommand{\href}[2]{#2}\begingroup\raggedright\endgroup

\end{document}